# Electronic multipoles in second harmonic generation and neutron scattering

G. van der Laan[1] and S. W. Lovesey[1,2],

[1]Diamond Light Source, Harwell Science and Innovation Campus, Didcot, Oxfordshire OX11 0DE, UK

[2]ISIS Facility, STFC, Didcot, Oxfordshire OX11 0QX, UK

**Abstract** Nonlinear optics, and particularly second harmonic generation (SHG), is increasingly used in many modern disciplines from material characterization in physical sciences to bio-imaging in medicine and optical signal processing in information technology. We present a theoretical analysis providing a strong estimate for the energy-integrated SHG response. Requirements of symmetry in time and space are fully respected in the calculation, and estimates of natural and magnetic circular dichroic signals are superior to previous ones. Like symmetry requirements are traced in the amplitude for magnetic neutron scattering, which includes all axial and polar (Dirac) contributions. Our method of working, in terms of spherical multipoles and implementation of symmetry, could be of use in a variety of other probes of electronic properties.

## I. INTRODUCTION

The study reported here contributes to the quest for a deeper understanding of intriguing, and potentially useful, electronic properties of substances. It focuses on what can be learnt from the response of chiral and magnetic materials to their illumination by beams of photons or neutrons. Following introductions to the two experimental techniques, suitable response functions are derived within an atomic framework. To this end, we exploit fundamental properties of observable response functions that also must apply to results inferred from other experimental techniques, and not just photon and neutron scattering that are of immediate interest. In our study, electronic degrees of freedom in the ground state are encapsulated in multipoles created from spherical tensor operators with specific discrete symmetries. This representation lends itself to a unification of results from a raft of experimental techniques.

We specialize the study to second harmonic generation and magnetic neutron scattering. They might seem unnatural bedfellows, for one technique is a mainstay in studies of non-magnetic electronic properties that violate spatial invariance and the other is the preferred technique to investigate conventional, axial magnetism expressed by spin and orbital angular momentum operators alone. However, a theory of magnetic neutron scattering by parity-odd multipoles has been ratified in diffraction by anapoles (Dirac dipoles) expressed by spin, angular momentum and spatial operators, while Dirac quadrupoles with the same operators have been shown to account for Bragg diffraction patterns gathered on many ceramic superconductors.

## A. Photon scattering

Nonlinear optical phenomena encompass physical processes originating from the interaction of light with matter, which modify the incoming electromagnetic field generating radiation of different frequency [1, 2, 3]. Such processes occur when a material interacts with intense light whereby its response yields fundamentally different properties than the one observed in the linear regime. Only after the introduction of the laser in 1960 [4] the first experimental evidence of nonlinear phenomena was reported in 1961 by Franken *et al.* [5]. They detected the frequency doubling of radiation passing through a nonlinear crystal; the process of second harmonic generation in the visible light region, a phenomenon previously known for radio waves only. This observation is regarded as the beginning of the field of nonlinear optics, although two-photon absorption had already been predicted by Maria Göppert-Mayer in 1931 [6]. Armstrong *et al.* [7] and Loudon *et al.* [8] treated nonlinear optics within the general framework of response theory in 1962.

The Kramers-Heisenberg dispersion formula uses the absorption of a photon and subsequent emission of a photon, i.e., it is a two-photon process and quadratic in the electric field strength **E**. A corresponding electronic response of the illuminated substance is polarization **P** linear in **E**. Within a perturbative development, nonlinear optical phenomena are classified according to the number of photons simultaneously involved, or, more strictly, the number of times **E** is engaged. In line with this development, the electronic polarization **P** is written in ascending powers of **E** (using the Einstein summation convention on Cartesian labels),

$$P_i = \epsilon_0[\chi_{ij}^{(1)}E_j + \chi_{ijk}^{(2)}E_jE_k + \chi_{ijkl}^{(3)}E_jE_kE_l + \ldots],$$

where the proportionality constant $\chi^{(n)}$ is known as the n-th order susceptibility, which is a (n + 1)-th rank tensor, and $\epsilon_0$ is the permittivity of free space.

The first-order susceptibility $\chi_{ij}^{(1)}$ refers to the linear optical response and the other terms are referred to as nonlinear . According to the von Neumann Principle the components $\chi^{(n)}$ of the nonlinear susceptibility are determined by the magneto-crystalline symmetry. For instance, second-order response is zero for materials with inversion symmetry for light that can be described within the electric-dipole approximation. On the contrary, third order effects are present.

Second harmonic generation (SHG) is one of the processes related to the second-order non-linear susceptibility $\chi^{(2)}$, where two incident photons of the same energy, interacting with matter, generate a third photon at twice the energy in a coherent manner as in Fig. 1. If the initial photon energies are different, the energy of the generated photon may be either the sum or the difference of the incident photon energies, which are known as sum frequency generation (SFG) and difference frequency generation, respectively. Similarly, the excitation of matter with a single impinging photon can trigger the emission of two output photons with a total energy equal to the exciting photons. This process is known as spontaneous parametric down conversion. The electro-optic (Kerr) effect is a change to the refractive index of a substance that is similarly a quadratic function of the electric field strength. Third-order nonlinear processes are related to $\chi^{(3)}$ and involve the interaction of four photons, giving rise to a variety of mixing processes, known as four wave mixing. The four-wave mixing process, where three incident photons have the same energy, generates a fourth trifold more energetic photon, is also known as third harmonic generation. As the initial and final quantum mechanical states of the system interacting with light are the same these processes are defined as parametric, contrary to processes where the system is left in an excited state such as stimulated Raman, Brillouin, and Rayleigh scattering. For the case of the (linear) refractive index the real and imaginary part describes a response from parametric and nonparametric processes, respectively. As no radiation is absorbed, a parametric process suffers less sample damage.

Higher-harmonic generation gives rise to a myriad of phenomena, such as spatial optical solitons, namely, a field that does not change during propagation because diffraction is balanced by the (nonlinear) optical Kerr effect. In the phenomenon of optical phase conjugation, the propagation direction and phase variation of a beam of light is exactly reversed. Two interacting beams simultaneously interact in a nonlinear optical material to form a real-time diffraction pattern, in the material. The third incident beam diffracts at this dynamic hologram, and, in the process, reads out the phase-conjugate wave.

In this paper, we are primarily concerned with SHG, but results can be extended to other nonlinear effects. A simple picture to describe the SHG process is to consider a three-level system depicted in Fig. 1. An incident photon excites an electron of the system, which is promoted to an empty state, and a second photon excites it to the next level. The state then de-excites to the equilibrium ground-state under emission of a photon that, due to energy conservation, has twice the energy and frequency of the original photons. However, SHG is not just a three-step process, but instead a single three-body interaction. The intermediate states can be thought as virtual states described by many-body wave functions, subjected to microscopic interactions such as the Coulomb and exchange interactions and the spin-orbit coupling.

A wide variety of applications of SHG have seen the light over the last decades. One of the most commonly used is frequency doubling in laser systems, letting the beam pass through a nonlinear crystal [9,10] with a large $\chi^{(2)}$ at the desired frequency. These properties are material dependent and much work has gone into the research of high-efficiency nonlinear materials [9,11].

An important application of SHG is as a probe for spectroscopy or microscopy. As SHG excited by electric-dipole radiation is forbidden for a centrosymmetric medium, it is consequently highly sensitive to symmetry breaking. This makes it a selective probe for surfaces and interfaces of centrosymmetric media, where the bulk does not contribute and the frequency-doubled signal is therefore characteristic of the first few atomic layers close to the surface or interface. It also allows a time-resolved in-situ monitoring of the surface reconstruction, of its chemistry, when molecules or other adsorbates are deposited [12]. This application ranges over a large variety of materials: metal surfaces, metal-electrolyte interfaces, semiconductors, oxides, insulator surfaces/interfaces, etc. [13,14]. SHG experiments have been frequently used to determine the average orientation of molecules adsorbed at surfaces, through measurements of the polarization dependence and phase [13]. In the last decade it has been affirmed as a selective non-destructive spectroscopy technique for the study of surfaces [15], superlattices [16-18], interfaces [19,20] and two-dimensional materials [21]. Furthermore, second harmonic imaging microscopy has been employed to the study and imaging of cells and biological membranes, and especially collagen. It is a non-destructive probe that allows to study *in-vivo* biological systems in their environment [22-24].

At the x-ray region optical wave mixing was proposed nearly half a century ago as an atomic-scale probe of light-matter interactions [25,26]. The recent advent of free-electron lasers (FELs) in the energy range from extreme ultraviolet to x rays allows us to explore these effects involving core-level resonances [27]. Yamamoto *et al.* [28] measured SHG at the Fe $3p$ edge of gallium ferrate ($GaFeO_3$) using soft x-ray FEL radiation. Other nonlinear optical techniques observed in the extreme-ultraviolet and x-ray region include SFG [29], four-wave mixing [30], and x-ray two-photon absorption [31]. The ultrafast x-ray pulses from a high brightness x-ray FEL provide the capability for time-resolved probing of atomic scale structure and electronic states in a material.

Similar to breaking of space-inversion symmetry, breaking of time-inversion symmetry by long-range magnetic ordering or an applied magnetic field leads to new contributions to SHG, which can be used to probe the magnetic structure [32,33]. Magnetic SHG is especially powerful in the case of antiferromagnets where, with a few specific exceptions, the usual magneto-optical methods fail because of the absence of bulk magnetization [34]. Rotation of the plane of polarization of reflected light (magneto-optic Kerr effect, often referred to by the acronym MOKE) is one direct manifestation of broken time-reversal symmetry. Despite the absence of bulk magnetization, magnetic crystal classes *m'm'm'*, 2/*m'* and 2*m'm'* generate polarization rotation according Orenstein [35]. For these magnetic symmetries, the Kerr effect is mediated by magnetoelectric coupling, which can arise when antiferromagnetic order breaks inversion symmetry.

Our calculation of the SHG response in an atomic framework set out in Section IV follows two sections that record necessary preliminary material about space and time symmetries. At the end, there is a strong estimate of the energy integrated SHG response and the corresponding natural and magnetic circular dichroic signals, i.e., a difference in absorption on reversing the polarity of circular polarization in the primary beam. Formal aspects of the calculation of dichroic signals from the SHG response are reviewed in §§ A.3 and A.6 of Ref. [6]. For present purposes, it is sufficient to note that a dichroic signal is proportional to the energy integrated SHG response.

**B. Neutron scattering**

A direct observation of neutron diffraction by anapoles (Dirac dipoles) [36, 37], and compelling evidence from neutron Bragg diffraction patterns that ceramic superconductors support Dirac (magnetoelectric) quadrupoles [38] are reasons enough to revisit theories of neutron scattering. The onset of magnetic Bragg diffraction in ceramic superconductors occurs at temperatures associated with the appearance of the pseudo-gap state as determined by probes such as ARPES, NMR, and optical conductivity [39]. By and large, theories of neutron scattering in widespread use simply rely on conventional electronic dipoles and, even then, ignore orbital magnetism. The latter cannot be ignored in a meaningful account of rare-earth magnetism, for example. Trammell [40] provided a theory of magnetic neutron scattering, and an updated version of his work in terms of axial (parity-even) multipoles is readily available [41, 42, 43]. A feature of Trammell's work is the restriction to atomic states in a $J$-manifold, whereupon all axial multipoles have an odd rank. This feature is thoughtlessly carried over to most reported theoretical interpretations of experimental results. One example is found in a simulation of scattering from cerium ions in a pyrochlore, while in possession of evidence that two $J$-manifolds are required to represent atomic $Ce^{3+}$ in the material [44]. In this case, with $J = 5/2$ and $J' = 7/2$, even rank axial multipoles are permitted, and they represent entanglement of a spin anapole with additional spatial degrees of freedom [43, 45]. Inclusion of Dirac multipoles, permitted for ions at sites in a magnetic material without a centre of spatial inversion symmetry, eventually completes a theory of neutron scattering by unpaired electrons.

Bragg diffraction experiments have demonstrated that underdoped $HgBa_2CuO_{4+\delta}$ (Hg1201) samples possess magnetic order indexed on the chemical structure, and the same is found in identical experiments on a structurally more complicated cuprate $YBa_2Cu_3O_{6+x}$ (YBCO), with two $CuO_2$ plaquettes in a unit cell [46, 47]. A ferro-type motif of Dirac quadrupoles depicted in Fig. 2 is consistent with all available neutron diffraction data [38]. The story for Hg1201 is similar to YBCO, where neutron magnetic Bragg diffraction and the Kerr effect have been observed and successfully related to Dirac quadrupoles [35]. While Hg1201 and YBCO possess the same magnetic space-group ($Cm'm'm'$, magnetic crystal-class $m'm'm'$), not surprisingly, there are significant differences in the symmetries at sites used by Cu ions. Notably, Cu ions in non-magnetic Hg1201 occupy sites that are centres of spatial inversion symmetry. However, condensation of Dirac quadrupoles in the pseudo-gap phase breaks the inversion symmetry.

Beginning in Section V and continuing in the Appendix, we scrutinize all contributions to the amplitude of magnetic neutron scattering against requirements of symmetry established in Section III. The orbital-spin contribution is relatively straightforward, which likely explains why usually it is the sole contribution to investigations. Yet it contains a frequently overlooked subtlety in the appearance of multipoles of even rank, which tell about correlations in spatial and anapole variables. The orbital contribution to the amplitude is complicated by comparison, and our study of it spills over to the Appendix. For one thing, it contains three types of radial integrals set against one in the orbital-spin contribution to scattering. Reduction to one type of integral in the entire scattering amplitude is a measure of the great simplification that occurs for the special case of equivalent electrons, e.g., electrons in the rare-earth 4f shell. Multipoles in the orbital contribution have an odd rank in all cases.

## II. PRELIMINARY REMARKS

Cartesian ($x$, $y$, $z$) and spherical components $R_Q$ of a vector **R** are related by $x = (R_{-1} - R_{+1})/\sqrt{2}$, $y = i(R_{-1} + R_{+1})/\sqrt{2}$, $z = R_0$. Evidently, spherical components of a dipole are complex for $Q$ different from zero, and $\{R_Q\}^\times = (-1)^Q R_{-Q}$, where $^\times$ is complex conjugation. Operators we use possess discrete symmetries with respect to the reversal of the coordinates of space and time, with corresponding signatures $\sigma_\pi$ and $\sigma_\theta$. An operator is parity-even (odd) for $\sigma_\pi = +1$ ($-1$), while time-even (odd) for $\sigma_\theta = +1$ ($-1$). Parity and time inversions are conjugate operations that reverse the sign of a four-vector. The vector (dipole) **R** is polar with $\sigma_\pi = -1$, whereas spin **S** (and orbital angular momentum **L** and **J** = **S** + **L**) is axial with $\sigma_\pi = +1$. We adopt the convention that $\sigma_\theta$ is the sign difference between time-reversal and complex conjugation. Since the two operations applied separately to $R_Q$ yield identical results the time signature $\sigma_\theta = +1$, while $S_Q$ possesses a time signature $\sigma_\theta = -1$. More generally, the rank of a spherical tensor operator is labelled by a positive integer $K$, and it possesses $(2K + 1)$ projections $-K \leq Q \leq K$. Our operators obey $[\mathcal{O}^K_Q]^+ = (-1)^Q \mathcal{O}^K_{-Q}$, where $^+$ denotes Hermitian conjugation, and the diagonal component $\mathcal{O}^K_0$ is Hermitian.

The action of our time-reversal operator $\theta$ on a time-dependent wavefunction of the system $\psi(\mathbf{R}, t)$ is $\theta\psi(\mathbf{R}, t) = \psi^\times(\mathbf{R}, -t)$, while $\theta\psi(\mathbf{H}) = \psi^\times(\mathbf{H}) = \psi(-\mathbf{H})$ for a stationary state subject to a magnetic field **H**. A matrix element of an arbitrary operator $B$ satisfies [48],

$$(\psi, B\varphi) = (\theta\varphi, \bar{B}^+\theta\psi), \qquad (1)$$

with $\bar{B} = \theta B \theta^{-1}$, which follows from standard properties $\theta(c\psi) = c^\times(\theta\psi)$ and $(\theta\psi, \theta B\varphi) = (\psi, B\varphi)^\times$ (c is a classical number). Whereupon,

$$(\psi, B\varphi)_\mathbf{H} = (\varphi, \bar{B}^+\psi)_{-\mathbf{H}} = \sigma_\theta (\varphi, B\psi)_{-\mathbf{H}}, \qquad (2)$$

and our definition $\bar{B} = \sigma_\theta B^+$ equates change in the polarity of an applied magnetic field with change in the mean value of $B$ with respect to the reversal of time, which is a sensible outcome.

The Wigner 3j-symbol and Clebsch-Gordan coefficient $(a\alpha\, b\beta|KQ)$ are related by,

$$(a\alpha\, b\beta|KQ) = (-1)^{-a+b-Q} \sqrt{2K+1} \begin{pmatrix} a & b & K \\ \alpha & \beta & -Q \end{pmatrix}. \qquad (3)$$

The 3j-symbol is invariant with respect to inversion of spatial coordinates [49, 50]. The parity operator commutes with the rotation operator, because rotations are generated by parity-even operators $J_\alpha$. Thus, states $|JM\rangle$ that differ only in projections $M$ have the same parity. The time-reversal operator $\theta$ reverses the sign of a projection. If we equate $\theta(c|JM\rangle)$ with $c^\times (-1)^{J-M} |J,-M\rangle$, a composite state,

$$|JM\rangle = \sum_{mm'} |jm\rangle |j'm'\rangle \, (jm\, j'm'|JM), \tag{4}$$

is fully compatible with it. Notably, $\theta^2|JM\rangle = (-1)^{2J}|JM\rangle = \pm|JM\rangle$, where the upper sign is for integer $J$ and the lower sign is for half-integer $J$.

The Wigner-Eckart Theorem says a matrix element $(JM, \mathcal{O}^K_Q\, J'M') \equiv \langle JM|\mathcal{O}^K_Q|J'M'\rangle$ is related to a reduced matrix element (RME) that does not depend on projections, $M$ and $M'$. We use a standard definition of the theorem whereby,

$$\langle JM|\mathcal{O}^K_Q|J'M'\rangle = (-1)^{J-M} \begin{pmatrix} J & K & J' \\ -M & Q & M' \end{pmatrix} (J\|\mathcal{O}^K\|J'). \tag{5}$$

The RME $(J\|\mathcal{O}^K\|J')$ is also called a double-barred matrix element. Total angular momentum $J = (l \pm \sigma)$, with $\sigma = 1/2$. Electronic matrix elements and RMEs are diagonal with respect to the magnitude of the spin, of course, and $\sigma$ is not made explicit henceforth.

### III. RESPONSE FUNCTIONS

Two key identities must be satisfied in the RME of a response function. The first is a straightforward consequence of the definition $[\mathcal{O}^K_Q]^+ = (-1)^Q \mathcal{O}^K_{-Q}$ and the Wigner-Eckart Theorem, namely (§§ A.3 and A.5 in Ref. [50]),

$$(lJ\|\mathcal{O}^K\|l'J') = (-1)^{J-J'} (l'J'\|\mathcal{O}^K\|lJ)^\times, \tag{6}$$

and it is worth noting that $(-1)^{J-J'} = (-1)^{J'-J}$. The second identity is a direct outcome of (1) for which we now use an operator of rank $K$ with parity signature $\sigma_\pi$, i.e., application of the parity operator results in $(P_\pi B^K P_\pi^{-1}) = \sigma_\pi B^K$. Simultaneous inversion of space and time coordinates on an elemental state $|JM\rangle$ is represented by,

$$P_\pi \theta(c|JM\rangle) = c^\times (-1)^{J-M} |J,-M\rangle, \tag{7}$$

whereupon [3],

$$(l'J'\|B^K\|lJ) = (-1)^{J'-J} (-1)^K \sigma_\pi \sigma_\theta (lJ\|B^K\|l'J'), \tag{8}$$

follows from the Wigner-Eckart Theorem. Combining (6) and (8),

$$(lJ\|\mathcal{O}^K\|l'J') = (-1)^K \sigma_\pi \sigma_\theta (lJ\|\mathcal{O}^K\|l'J')^\times. \tag{9}$$

We turn to applications of (6) and (8).

First, a useful result for the RME of a response function. It is simple to verify that,

$$(lJ||O^K||l'J') = \chi [Z^K(lJ, l'J') + \rho Z^K(l'J', lJ)], \tag{10}$$

obeys (6) and (8) for purely real $Z^K(lJ, l'J')$. Phase factors are found to be $\chi^2 = \{(-1)^K \sigma_\pi \sigma_\theta\}$ and $\rho = \chi^2 (-1)^{J'-J}$, meaning $\rho = \pm 1$ and $\chi$ purely real or imaginary.

### IV. SECOND HARMONIC GENERATION RESPONSE

By way of an introduction to SHG, the amplitude of x-ray scattering enhanced by an electric dipole - electric dipole (E1-E1) event derived from the Kramers-Heisenberg dispersion formula is a familiar response function. It is proportional to $\{\langle lJM|\boldsymbol{\varepsilon}'\cdot\mathbf{R}|\lambda jm\rangle \langle \lambda jm|\boldsymbol{\varepsilon}\cdot\mathbf{R}|lJ'M'\rangle\}$, where $\boldsymbol{\varepsilon}'$ ($\boldsymbol{\varepsilon}$) is the purely real photon polarization vector for the secondary (primary) absorption process, and $\mathbf{R}$ is the electronic dipole operator. Virtual intermediate states have atomic quantum numbers $\lambda jm$. If these are treated as spherically symmetric the dyadic can be replaced by its value integrated over projections $m$ in an application of the formula. Concomitant with neglect of angular anisotropy, energies of the intermediate states also are taken to be weak functions of the projections, and they will not contribute to our energy integrated signal. One finds [41],

$$\Sigma_m\{\langle lJM|\boldsymbol{\varepsilon}'\cdot\mathbf{R}|\lambda jm\rangle \langle \lambda jm|\boldsymbol{\varepsilon}\cdot\mathbf{R}|lJ'M'\rangle\} \propto \Sigma_{K,Q} (-1)^Q X^K_{-Q} \langle JM|T^K_Q|J'M'\rangle, \tag{11}$$

where $\mathbf{X}^K$ is a sole function of polarization vectors, and $\langle JM|T^K_Q|J'M'\rangle$ derived from dipole operators satisfies Eq. (5), i.e., $\mathbf{T}^K$ is an electronic spherical tensor in which the intermediate orbital angular momentum $\lambda$ is a parameter.

Components of $\mathbf{X}^K$ are evaluated from Eq. (4) after setting $j = j' = 1$ to represent simple vectors, while the Clebsch-Gordan coefficient embodies the triangle rule for addition of vectors to form spherical tensors $K = 0$, 1 and 2; $X^0_0 = -(1/\sqrt{3})(\boldsymbol{\varepsilon}'\cdot\boldsymbol{\varepsilon})$, $\mathbf{X}^1 = \{(i/\sqrt{2})(\boldsymbol{\varepsilon}'\times\boldsymbol{\varepsilon})\}$, and the traceless quadrupole $\mathbf{X}^2$ has a diagonal component $X^2_0 = \{(1/\sqrt{6})(3\varepsilon'_z\varepsilon_z - \boldsymbol{\varepsilon}'\cdot\boldsymbol{\varepsilon})\}$. The RME for $\mathbf{T}^K$ is purely real and thus identified with $Z(lJ, lJ')$ in Eq. (10). One finds $(lJ||T^K||lJ')$ is a sum on $a$ (spin) and $b$ (orbital) of a standard unit tensor $W^{(a,b)K}(lJ, lJ')$ restricted by the condition $(a + b + K)$ even. The unit tensor for one electron is defined in (A8). The simple result $W^{(a,b)K}(lJ, lJ') = (-1)^{J'-J} W^{(a,b)K}(lJ', lJ)$ is a direct consequence of said condition. Therefore, the RME in Eq. (10) is proportional to $\chi [1 + \chi^2]$, and $\chi^2 = \{(-1)^K \sigma_\theta\}$ for a parity-even event. In conclusion, the electronic multipole for an E1-E1 absorption event is different from zero when $(-1)^K = \sigma_\theta$, meaning the dipole is time-odd (magnetic) and the monopole and quadrupole are time-even (charge-like). Identical results can be derived from crossing-symmetry [41]. The results might be inferred from the observation that $\mathbf{X}^0$ and $\mathbf{X}^2$ differ in phase from $\mathbf{X}^1$ by 90°.

Electric quadrupole - electric quadrupole (E2-E2) enhancement is similarly described by Eq. (11) on replacing the dipole operator therein by $[(\boldsymbol{\varepsilon}\cdot\mathbf{R})(\mathbf{q}\cdot\mathbf{R})]$ where $\mathbf{q}$ is the primary photon wavevector, and likewise its secondary equivalent ($\boldsymbol{\varepsilon}\cdot\mathbf{q} = \boldsymbol{\varepsilon}'\cdot\mathbf{q}' = 0$). According to the triangle rule for addition of two tensors of rank two, E2-E2 scattering is described by five electronic multipoles $K = 0, ..., 4$. Note surprisingly, their time signature $\sigma_\theta = (-1)^K$. Beyond electric parity-even absorption considered thus far is parity-odd E1-E2 and, also, electric dipole - magnetic dipole (E1-M1) events [19]. Scattering amplitudes for these events are sums of multipoles that are time-even ($\sigma_\pi \sigma_\theta = -1$) and time-odd ($\sigma_\pi \sigma_\theta = +1$). Using dipoles by way of illustrations, these are expectation values of a displacement $\langle R_Q \rangle$ and an anapole (toroidal moment), respectively.

*SHG response* A Cartesian form of NCD from the SHG response uses an E1'-E1-E1 event. Our coordinates are defined by the primary beam parallel to the $z$-axis and σ-polarization parallel to x-axis. Secondary beam, engaged in an E1 event, is inclined to the $z$-axis and its polarization vector has a component parallel to it. The generic form of the energy integrated NCD signal is ([51] and § A.6 of Ref. [52]),

$$F(\text{NCD}) \propto P_2 \sum_{f,f'} \{ \langle g|x|f \rangle \langle f|y|f' \rangle \langle f'|z|g \rangle - \langle g|y|f \rangle \langle f|x|f' \rangle \langle f'|z|g \rangle \}. \tag{12}$$

Here, g denotes the electronic ground state and f and f' label intermediate states, while $P_2$ is a pseudo-scalar for helicity in the primary radiation. We proceed as in (11) and integrate out intermediate projections of the orbital angular momentum, with the same caveat about our energy integrated signal not being a function of intermediate energies, i.e., they are independent of labels f and f', to a good approximation. Again, products of matrix elements factor, as depicted in Fig. 3, although the analogue of the photon tensor $\mathbf{X}^K$ is more complicated [52]. Our goal is to produce operator equivalents for $F(\text{NCD})$ and $F(\text{MCD})$ defined by corresponding response RMEs that we derive in accord with Eq. (10).

Specifically, a result for $Z^K(lJ, l'J')$ in Eq. (10) suitable for NCD and MCD in the SHG response is calculated from $\{\langle lJM|\text{E1}|\lambda jm\rangle \langle \lambda jm|\text{E}p|\lambda'j'm'\rangle \langle \lambda'j'm'|\text{E1}|l'J'M'\rangle\}$, where index $p = 1, 2, \& 3$ labels modes available in the primary beam. Previously, we established that $p = 1$ (2) for NCD (MCD) [6]. Following sums on projections $m$ and $m'$ (Eq. (11) in Ref. [52]),

$$Z^K(lJ, l'J') = (-1)^{J-j} (lJ\|\mathbf{R}\|\lambda j) (\lambda j\|\mathbf{C}^p(\mathbf{R})\|\lambda'j') (\lambda'j'\|\mathbf{R}\|l'J')$$

$$\times \begin{Bmatrix} 1 & p & K \\ J' & J & j \end{Bmatrix} \begin{Bmatrix} p & 1 & p \\ J' & j & j' \end{Bmatrix}, \tag{13}$$

with the normalized spherical harmonic $\mathbf{C}^1(\mathbf{R}) = \mathbf{R}$. (Strictly speaking, the spatial argument is the unit vector $\mathbf{n} = \mathbf{R}/R$, whereas our use of $\mathbf{R}$ for the dipole operator might help visual tracking of the development.) The corresponding RME is purely real with $(lj\|\mathbf{C}^k\|l'j') = (-1)^{j-j'}(l'j'\|\mathbf{C}^k\|lj)$ in the general case, and the derivation of $Z^K(l'J', lJ)$ from Eq. (13) is then straightforward. After a re-coupling of variables in (13), the RME for the SHG response, defined by Eqs. (10) and (13), has an appealing form,

$$(lJ\|\mathcal{O}^K\|l'J') = \chi (-1)^{J-j} \sum_x (2x+1) \begin{Bmatrix} 1 & 1 & x \\ p & K & p \end{Bmatrix} \begin{Bmatrix} 1 & 1 & x \\ j & j' & p \\ J & J' & K \end{Bmatrix} [1 + \sigma_\pi \sigma_\theta (-1)^{x+p}]$$

$$\times (lJ \|\mathbf{R}\|\lambda j) \, (\lambda j\|\mathbf{C}^p(\mathbf{R})\|\lambda'j') \, (\lambda'j'\|\mathbf{R}\|l'J'). \tag{14}$$

Both dichroic signals of interest have $\sigma_\pi \sigma_\theta = -1$, and the index $x = 0, 1, 2$ in Eq. (14) is uniquely defined, namely, $x = 2$ for NCD ($p = 1$, $K = 2$) and $x = 1$ for MCD ($p = 2$, $K$ odd).

In order to perform sums on $j$ and $j'$ in Eq. (14) we need an explicit result for the RME of a spherical harmonic $(lj\|\mathbf{C}^k\|l'j')$. In our chosen s-l coupling scheme [50],

$$(\sigma lj\|\mathbf{C}^k\|\sigma l'j') = (-1)^{\sigma + l' + j + k} [(2j + 1)(2j' + 1)]^{1/2} \begin{Bmatrix} l & j & \sigma \\ j' & l' & k \end{Bmatrix} (l\|\mathbf{C}^k\|l'), \tag{15}$$

with,

$$(l\|\mathbf{C}^k\|l') = (-1)^l [(2l + 1)(2l' + 1)]^{1/2} \begin{pmatrix} l & k & l' \\ 0 & 0 & 0 \end{pmatrix},$$

and the 3j-symbol can be different from zero for $(l + k + l')$ even.

We define the response RME $(l\|\mathcal{O}^K\|l')$ to conform with an identical s-l coupling scheme, i.e., $(lJ\|\mathcal{O}^K\|l'J')$ and $(l\|\mathcal{O}^K\|l')$ are related as in Eq. (15), with,

$$(l\|\mathcal{O}^K\|l') = -\chi \sum_x (2x + 1) \begin{Bmatrix} 1 & 1 & x \\ p & K & p \end{Bmatrix} \begin{Bmatrix} l & l' & K \\ 1 & 1 & x \\ \lambda & \lambda' & p \end{Bmatrix} [1 + \sigma_\pi \sigma_\theta (-1)^{x + p}]$$

$$\times (l\|\mathbf{R}\|\lambda) \, (\lambda\|\mathbf{C}^p(\mathbf{R})\|\lambda') \, (\lambda'\|\mathbf{R}\|l'). \tag{16}$$

Eqs. (14) and (16) are core results for the SHG response.

Intermediate angular momenta $\lambda$ and $\lambda'$ in Eq. (16) can be integrated out for the NCD signal, with the result,

$$(l\|\mathcal{O}^2(\text{NCD})\|l') = -\chi \sqrt{(2/15)} \, (l\|\{\mathbf{R} \otimes \mathbf{C}^2\}^2\|l'), \tag{17}$$

and $\chi^2 = -1$. Eq. (17) uses a standard definition of a tensor product modelled on Eq. (4),

$$\{\mathbf{A}^a \otimes \mathbf{B}^b\}^K_Q = \sum_{\alpha,\beta} \mathbf{A}^a_\alpha \mathbf{B}^b_\beta \, (a\alpha \, b\beta | KQ). \tag{18}$$

Eq. (17) says that the NCD signal in the SHG response can be calculated using matrix elements of the operator $\{\mathbf{R} \otimes \mathbf{C}^2\}^2$, which is evidently parity-odd and time-even. As in our previous calculation, the operator equivalent is a dyadic [52]. The corresponding MCD result is,

$$(l\|\mathcal{O}^K(\text{MCD})\|l') = -\chi [6/(2K + 1)]^{1/2} \begin{Bmatrix} 1 & 1 & 1 \\ 2 & K & 2 \end{Bmatrix} (l\|\{\mathbf{C}^2 \otimes \mathbf{L}\}^K\|l'), \tag{19}$$

with $K = 1$ and 3, and $\chi^2 = +1$. The operator equivalent $\{\mathbf{C}^2 \otimes \mathbf{L}\}^K$ is manifestly both parity-even and time-odd. Multipoles calculated with Eq. (19) are used in the MCD signal $F(\text{MCD}) = \{q_0 \, P_2 \, [\sqrt{2}\langle \mathcal{O}^1{}_0 \rangle + \sqrt{3}\langle \mathcal{O}^3{}_0 \rangle]\}$, where $q_0 \equiv q_z$ represents the primary wavevector [52].

## V. NEUTRON SCATTERING

The amplitude for magnetic scattering of neutrons is written $\mathbf{Q}_\perp = [\boldsymbol{\kappa} \times (\mathbf{Q} \times \boldsymbol{\kappa})]$ using a unit wave vector $\boldsymbol{\kappa} = \mathbf{k}/k$ where $\mathbf{k}$ is the scattering wavevector. An intermediate operator can be written,

$$\mathbf{Q} = \exp(i\mathbf{R}_j \bullet \mathbf{k}) \, [\mathbf{s}_j - (i/\hbar k)(\boldsymbol{\kappa} \times \mathbf{p}_j)], \tag{20}$$

in which $\mathbf{R}_j$, $\mathbf{s}_j$ and $\mathbf{p}_j$ are operators for position, spin and linear momentum of unpaired electrons, respectively. Note that $\mathbf{Q}$ is arbitrary to within any function proportional to $\boldsymbol{\kappa}$.

*Orbital-spin contribution* If $\mathbf{k}$ is treated as a small parameter,

$$\{\exp(i\mathbf{R} \bullet \mathbf{k}) \, \mathbf{s}\} = \mathbf{s} + i(\mathbf{R} \bullet \mathbf{k}) \, \mathbf{s} - \ldots \tag{21}$$

The leading term contributes to an over-used approximation for $\mathbf{Q}$, namely, $\mathbf{Q} \approx (\mathbf{L} + 2\mathbf{S})/2$ for an isolated ion with orbital angular momentum. As for the second, manifestly parity-odd term in Eq. (21) there is a standard decomposition,

$$i(\mathbf{R} \bullet \mathbf{k}) \, \mathbf{s}_\alpha = (kR) \, i\kappa_\beta \{(1/3) \, \delta_{\alpha\beta} \, \mathbf{s} \bullet \mathbf{n} + (1/2) \, \varepsilon_{\alpha\beta\gamma} \, (\mathbf{s} \times \mathbf{n})_\gamma$$

$$+ (1/2) \, [\mathbf{s}_\alpha \mathbf{n}_\beta + \mathbf{s}_\beta \mathbf{n}_\alpha - (2/3) \, \delta_{\alpha\beta} \, \mathbf{s} \bullet \mathbf{n}]\}, \tag{22}$$

where $\mathbf{n} = \mathbf{R}/R$ (Einstein summation convention). The scalar operator does not contribute to $\mathbf{Q}_\perp$, i.e., the electronic Dirac monopole is not visible in neutron scattering, although it is visible in resonance enhance x-ray scattering [41]. Next in line, the spin anapole contribution $\{\kappa_\beta \, \varepsilon_{\alpha\beta\gamma} \, (\mathbf{s} \times \mathbf{n})_\gamma\} = [\boldsymbol{\kappa} \times (\mathbf{s} \times \mathbf{n})]_\alpha$ has been unambiguously detected in diffraction by two compounds with the C15 cubic Laves structure [36, 37]. Lastly, the traceless Dirac quadrupole accounts for Bragg diffraction patterns collected on pseudo-gap phases of Hg1201 and YBCO [38].

For arbitrary magnitudes of $\mathbf{k}$ we make use of a purely real function ($h_K$) that is the mean value of the spherical Bessel function $j_K(kR)$ with respect to radial parts of atomic orbitals. It reduces to a standard radial integral $\langle j_K(k) \rangle$ for equivalent electrons [53]; Fig. 4 contains ($h_1$) and $\langle j_0(k) \rangle$ for samarium, and definitions are derived from Eq. (A4). Integer $K$ is even for equivalent electrons, and odd for orbitals with opposing parities. We go on to find,

$$\{\exp(i\mathbf{R} \bullet \mathbf{k}) \, \mathbf{s}\} = \sum_{K, K'} i^{K'+1} (2K + 1)^{1/2} \, \{\mathbf{C}^K(\boldsymbol{\kappa}) \otimes \mathbf{H}^{K'}\}^1. \tag{23}$$

The RME of the tensor operator,

$$\mathbf{H}^{K'} = (-i)^{K'+K+1} [(2K + 1)(2K' + 1)/3]^{1/2} (h_K) \{\mathbf{s} \otimes \mathbf{C}^K(\mathbf{n})\}^{K'}, \qquad (24)$$

satisfies the fundamental requirement Eq. (10). To this end, $(lJ\|\{\mathbf{s} \otimes \mathbf{C}^K(\mathbf{n})\}^{K'}\|l'J') = \{\sqrt{(3/2)} (l\|\mathbf{C}^K\|l') W^{(1,K)K'}(lJ, l'J')\}$ with $(K + l + l')$ even, $K' = K, K \pm 1$ and a maximum rank $K' = (1 + l + l')$. Eq. (22) is recovered from (23) and (24) using $j_1(kR) \approx (kR)/3$ for a small argument, and $K' = 1$ and 2. Values of the unit tensor for equivalent electrons are tabulated [43]. The rank $K'$ is odd if electrons belong to a manifold $J = J'$, which can be verified from (A8) for a single electron and states $l = l'$.

*Orbital contribution* An analogue of the spin dipole in (21),

$$-(i/\hbar k) \{\exp(i\mathbf{R} \cdot \mathbf{k})(\boldsymbol{\kappa} \times \mathbf{p})\} = \{(1/2)[\langle j_0(k)\rangle + \langle j_2(k)\rangle] \mathbf{L} + i(\boldsymbol{\kappa} \times \mathbf{D})\} + \ldots, \qquad (25)$$

uses a Dirac dipole $\mathbf{D} = (1/2)[i(g_1)\mathbf{n} - (j_0)\boldsymbol{\Omega}]$, where $\boldsymbol{\Omega}$ is an orbital anapole defined in (A5), and integrals $(j_0)$ and $(g_1)$ are made with radial functions belonging to states with opposing parities, e.g., atomic 4f and 5d states. Unlike (21), Eq. (25) is valid for arbitrary values of the wave vector. While $\langle j_0(0)\rangle = 1$ for equivalent electrons, $(j_0)$ and $(g_1)$, depicted in Fig. 4, both diverge as $k$ approaches zero.

An RME for the orbital operator (25) that includes all tensor operators is complicated, as illustrated by the comparatively simple result for equivalent electrons,

$$-\{\exp(i\mathbf{R} \cdot \mathbf{k})(\boldsymbol{\kappa} \times \boldsymbol{\nabla})\} = \sum_{K'} \{\mathbf{C}^{K'-1}(\boldsymbol{\kappa}) \otimes \mathbf{O}^{K'}\}^1, \qquad (26)$$

with odd $K' = 1, 3, \ldots (2l - 1)$, and a tensor operator defined by the RME,

$$(l\|\mathbf{O}^{K'}\|l) = k\, A(K', l)\, [\langle j_{K'-1}(k)\rangle + \langle j_{K'+1}(k)\rangle], \qquad (27)$$

where the purely real function,

$$A(K', l) = i^{K'-1} [K'(K'+1) - 2(l+1)]^{-1} (l\|\{\mathbf{L} \otimes \mathbf{C}^{K'}(\mathbf{n})\}^{K'}\|l+1)$$

$$\times \{[K'(2K'-1)(2K'+1)(2l+1)(2l+2+K')(2l+1-K')]/[3(2l+3)]\}^{1/2}, \qquad (28)$$

where we use definition (A3). The RME $(lJ\|\mathbf{O}^{K'}\|lJ') = \{\sqrt{2}(l\|\mathbf{O}^{K'}\|l) W^{(0,K')K'}(lJ, lJ')\}$ complies with Eqs. (6) and (8), given $\sigma_\pi \sigma_\theta = -1$ ($\sigma_\theta = -1$, $\sigma_\pi = +1$,) and $K'$ odd. One finds $A(1, l) = (l\|\mathbf{L}\|l)/2$, consistent with (25). Results (26) - (28) are precisely Eq. (11.48) in Ref. [54].

Contributions to the Dirac dipole $\mathbf{D}$, and all higher-order operators are discovered in an expression suitable for atomic states with opposing parities $l \neq l'$. It is delegated to an Appendix on account of its complexity. One thing to note here is that the result (A2) is an explicit example of the response RME in Eq. (10).

## VI. DISCUSSION

Optical radiation can be expanded in multipole terms, such as electric dipole (E1), electric quadrupole (E2), magnetic dipole (M1), together with higher-order terms that can normally be neglected. In a two-photon process, described by the Kramer-Heisenberg formulism, the optical activity induced by a pure transition must have even parity, so that only interference terms can give odd parity. The pure E1-E1 process produces magnetic circular dichroism (MCD), which is a time-odd, parity-even event. The E1-M1 interference in the visible region allows natural circular dichroism (NCD) or optical rotation in powdered samples or in solution and single crystals. In the case of x-ray absorption, which involves core to valence shell excitations, M1 transitions are forbidden due to the restriction imposed by the monopole selection rule for the radial part. Although in the visible region the E2 transitions are negligibly small, their magnitude increases with photon energy, so that for harder x-rays the E1-E2 interference becomes significant, giving rise to NCD.

In this paper, we extended our inquiry to the second harmonic generation (SHG) response, with two primary photons of frequency $\omega$ and one secondary photon of frequency $2\omega$, as depicted in Fig. 1. Instead of the Kramer-Heisenberg formulism, SHG is adequately described by third-order perturbation theory, cf. Eq. (12). Using theoretical techniques from atomic physics we derived explicit expressions for electronic multipoles of the SHG response. Polarization-dependent photon spectroscopy (dichroism) of the SHG response is shown to reveal chiral and magnetic properties of a sample. Two dichroic signals are allowed with electric-dipole (E1) and electric-quadrupole (E2) scattering events, and both require circular polarization in the primary beam. NCD is derived from an (E1′-E1-E1) event, while no such dichroism arises from parity-odd events using magnetic dipoles (E1′-M1-M1) (Appendix B in Ref. [52]). A parity-even event (E1′-E2-E1) yields MCD [52].

Results Eqs. (17) and (19) for the SHG response are energy integrated signals. That is to say, from Eq. (17), the expectation value $\langle \{ \mathbf{R} \otimes \mathbf{C}^2(\mathbf{R}) \}^2 \rangle$ is the total NCD signal available from a substance in a suitably designed measurement, with a like statement derived from Eq. (19) for the MCD signal. Here, $\{ \, . \otimes . \, \}^K$ denotes a tensor product of rank $K$, $\mathbf{C}^2(\mathbf{R})$ is a spatial spherical harmonic of rank 2 normalized such that $\mathbf{C}^1(\mathbf{R}) = \mathbf{R}$, and the MCD signal is a sum of expectation values $\langle \{ \mathbf{C}^2(\mathbf{R}) \otimes \mathbf{L} \}^K \rangle$ with $K = 1$ and 3. A meaningful context for our results is the analogy between Eqs. (17) and (19) with celebrated sum-rules for conventional, parity-even dichroic signals [55-57] and their extensions to parity-odd signals [58], derived from the Kramers-Heisenberg dispersion formula. In the present setting, Eqs. (17) and (19) are products of the statement in Eq. (10) for the reduced matrix element of a response function. While fully compatible with earlier results for the same quantities [6], we submit that they are stronger statements. This claim is grounded on the re-coupling Eq. (14) of the exact result Eq. (13) for matrix elements in an E1′-E$p$-E1 event, depicted in Fig. 1, that was not previously accomplished [52]. A straightforward passage to Eqs. (17) and (19) is testament to insight from the re-coupling. Integrating out intermediate total angular momenta in the passage is a simplification that might not be called for, however, with Eq. (14) yielding superior estimates of available signals.

Turning to the second experimental technique in the present study, the orbital contribution to the magnetic neutron scattering amplitude can be calculated with Eq. (27), which has not been previously published, to the best of our knowledge. The equation, combined with (28), finds immediate application for a single electron in an atomic shell, e.g., $Ce^{3+}$ [45]. An extension of (27) to equivalent electrons is well-established and, therefore, not pursued here [43, 54]. The generalization to electrons in different atomic orbitals given in the Appendix is a specific example of the statement in Eq. (10) for the reduced matrix element of a response function. The orbital-spin contribution to the magnetic scattering amplitude Eq. (24), by comparison, is a much simpler response function that obeys Eq. (10). Perhaps deceptively simple, since Eq. (24) is the sole source of even rank multipoles in the scattering amplitude that are permitted by a correlation of the spin anapole and spatial degrees of freedom [43, 59].

## ACKNOWLEDGEMENT

Figure 2 was prepared by Dr D. D. Khalyavin.

## APPENDIX

Referring to Eq. (25), it is natural to separate the electronic linear momentum operator $\mathbf{p} = -i\hbar\nabla$ into its angular and radial components that we label by the letters $a$ and $r$. It can be shown that [60],

$$- (1/\hbar k) \exp(i\mathbf{R} \cdot \mathbf{k}) (\boldsymbol{\kappa} \times \nabla)_q = \sum_{K,K'} \{\mathbf{C}^K(\boldsymbol{\kappa}) \otimes \mathbf{O}^{K'}(K; a)\}^1_q \qquad (A1)$$

$$+ \sum_{K'} \{\mathbf{C}^{K'}(\boldsymbol{\kappa}) \otimes \mathbf{O}^{K'}(r)\}^1_q.$$

The radial tensor operator vanishes for equivalent electrons, as we will see, unlike the angular tensor (27) for $l = l'$. The angular tensor in (A1) necessarily depends on two indices, $K$ and $K'$, while a single index (rank) suffices for the radial tensor.

*Angular tensor*

$$\mathbf{O}^{K'}(K; a) = i^{K+K'-1} (2K+1)\sqrt{[3(2K'+1)]} \sum_{x,y} (-1)^{(1+K+x)/2} (j_x) (2x+1)(2y+1)$$

$$\times \begin{pmatrix} 1 & K & x \\ 0 & 0 & 0 \end{pmatrix} \begin{pmatrix} 1 & x & y \\ 0 & 0 & 0 \end{pmatrix} \begin{Bmatrix} 1 & x & y \\ K' & 1 & 1 \end{Bmatrix} \begin{Bmatrix} K & K' & 1 \\ 1 & 1 & x \end{Bmatrix} \{\mathbf{L} \otimes \mathbf{C}^y(\mathbf{n}) - (-1)^{K'+y} \mathbf{C}^y(\mathbf{n}) \otimes \mathbf{L}\}^{K'}, \quad (A2)$$

with $\mathbf{n} = \mathbf{R}/R$. In (A2),

$$(l'||\{\mathbf{L} \otimes \mathbf{C}^y(\mathbf{n})\}^{K'}||l) = (-1)^{l+l'} (l||\{\mathbf{C}^y(\mathbf{n}) \otimes \mathbf{L}\}^{K'}||l')$$

$$= (-1)^{K'+l+l'} \sqrt{(2K'+1)} \, (l'||\mathbf{L}||l') \, (l'||\mathbf{C}^y||l) \begin{Bmatrix} l' & l & K' \\ y & 1 & l' \end{Bmatrix}. \quad (A3)$$

The first thing to note is that $(l||\mathbf{O}^{K'}(K; a)||l') = (-1)^{K'+1} (l'||\mathbf{O}^{K'}(K; a)||l)$ follows from (A2) and $(y + l + l')$ even. In consequence, the RME satisfies Eq. (10). The radial integral is,

$$(j_x) = \int_0^\infty dR \, R^2 \, f_l(R) \, f_{l'}(R) \, \{j_x(kR)/kR\}, \quad (A4)$$

with $j_x(kR)$ a spherical Bessel function, and $f_l(R)$ and $f_{l'}(R)$ radial parts of electron orbitals labelled by their angular momenta $l$ and $l'$.

The rank $y$ defines the parity of the tensor operator, i.e., $\sigma_\pi = (-1)^y$. It is even for equivalent electrons in an atomic shell, $l = l'$, and $\mathbf{O}^{K'}(K; a)$ together with (A1) are identical to Eqs. (26) - (28). However, it is beholden to add a few words about properties of (A2) for equivalent electrons and recovery of the dipole contribution displayed in (25). Use of (A3) in (A2) shows that the tensor operator is different from zero for $(y + K')$ odd, meaning $K'$ odd. Moreover, $x = K'$ on inspection of (A2). The sum on $K$ in (A1) is removed by addition of a function proportional to $\kappa_q$ that leaves the amplitude for magnetic scattering of neutrons $\mathbf{Q}_\perp$ unchanged. The outcome is to set $K = K' - 1$ in (A1), and simultaneously replace all dependence on $K$ by $\sqrt{\{(2K'-1)/[3(K'+1)]\}}$ in (A2). It remains to perform the sum on $y = K' \pm 1$, and it is proportional to $(l||\mathbf{L}||l)$ for $K' = x = 1$ with a constant of proportionality that reproduces (25). The combination of radial integrals seen in (27) follows from the identity $j_n(z) = \{z \, [j_{n-1}(z) + j_{n+1}(z)]/(2n+1)\}$.

Consideration of orbitals with angular momenta that differ by an odd integer equates to parity $y$ odd. It follows that $x$ is even and $K$ is odd in (A2). One finds $y = K = K'$ for $K'$ odd, while $x = K'$ for $K'$ even. To recover the Dirac dipole $\mathbf{D} = -\{(1/2) \, (j_0) \, \mathbf{\Omega}\}$ contribution to (25), set $K' = 1$ and $y = K = 1$. Thereafter, two uses of an identity for the tensor product of two dipoles. First $\{\mathbf{C}^1(\kappa) \otimes \mathbf{O}^1\}^1 = (i/\sqrt{2}) \, (\kappa \times \mathbf{O}^1)$ in (A1), and the same identity yields,

$$\{\mathbf{L} \otimes \mathbf{C}^1(\mathbf{n}) - \mathbf{C}^1(\mathbf{n}) \otimes \mathbf{L}\}^1 = (i/\sqrt{2}) \, [\mathbf{L} \times \mathbf{n} - \mathbf{n} \times \mathbf{L}] = (i/\sqrt{2}) \, \mathbf{\Omega}. \quad (A5)$$

Properties of the orbital anapole $\mathbf{\Omega}$ are reviewed in Ref. [41].

It remains to define the radial tensor operator in (A1) and survey its properties. The radial integral in the result,

$$(l||\mathbf{O}^{K'}(r)||l') = i^{K'} (1/2) \, (g_{K'})^{l,\, l'} \, [K' (K'+1) (2K'+1)/3]^{1/2} \, (l||\mathbf{C}^{K'}(\mathbf{n})||l'), \quad (A6)$$

is odd with respect to the exchange of $l$ and $l'$. Specifically,

$$(g_{K'})^{l,\,l'} = (2K' + 1) \int_0^\infty dR R^2 [f_l(R) (d/dR)f_{l'}(R) \tag{A7}$$

$$- f_{l'}(R)(d/dR)f_l(R)]\{j_{K'}(kR)/Rk^2\},$$

and $(g_1)^{l,\,l'}$ is included in Fig. 4. The RME in (A6) satisfies Eqs. (6) and (8). Eq. (A6) evaluated for $K' = 1$, using $\mathbf{C}^1(\mathbf{n}) = \mathbf{n}$, yields the contribution made to the Dirac dipole. Note that the contribution to $\mathbf{D}$ uses the anti-Hermitian, time-odd and parity-odd operator $\{i\,\mathbf{n}\}$. The RME of the momentum operator is,

$$(l\|\mathbf{p}\|l') = (\hbar/2) [(l\|\boldsymbol{\Omega}\|l') - 2i(l\|\mathbf{n}\|l')].$$

We can infer from this result that the anapole and $\{i\,\mathbf{n}\}$ possess identical discrete symmetries.

The unit tensor for one electron [50],

$$W^{(a,b)K'}(\sigma l j, \sigma l' j') = [(2j + 1)(2K' + 1)(2j' + 1)]^{1/2} \begin{Bmatrix} \sigma & \sigma & a \\ l & l' & b \\ j & j' & K' \end{Bmatrix}, \tag{A8}$$

with $\sigma = 1/2$. The magnitude of the 9j-symbol is unchanged by an even or odd exchange of columns or rows, but an odd exchange changes its sign by a factor $(-1)^{\mathfrak{R}}$ with $\mathfrak{R} = (1 + a + l + l' + b + j + j' + K')$ [49, 50].

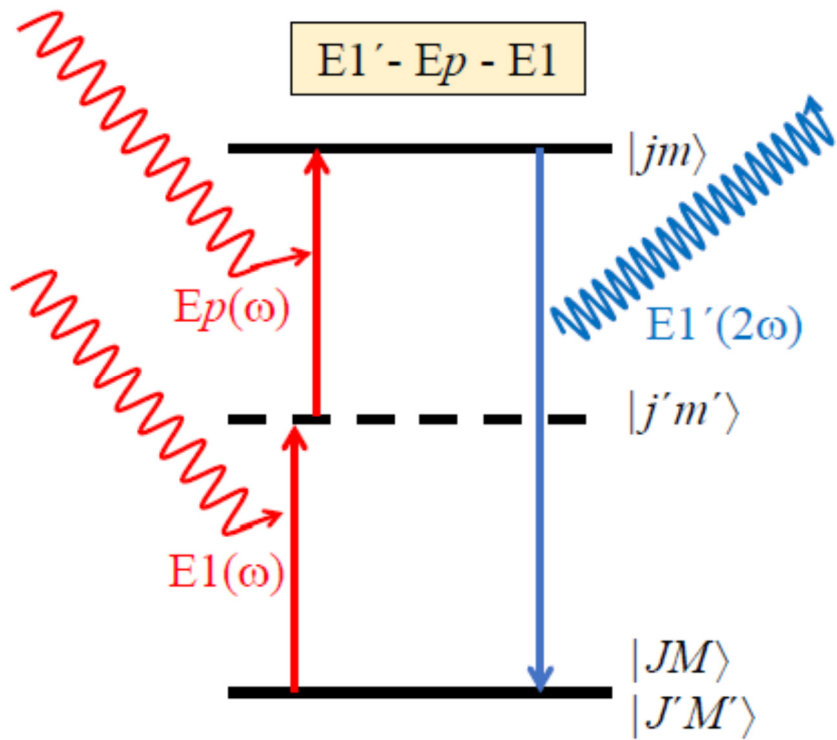

**Fig. 1.** Transitions that occur in second harmonic generation (frequency doubling) using three electronic operators E1′-E$p$-E1, where E$p$ = E1 or E2, and the primed operator E1′ relates to the secondary transition. Primary energy E = $\hbar\omega$ and integer $p$ labels modes available in the primary beam. Labelling of initial and final atomic states $|J'M'\rangle$, $|JM\rangle$ is used in Eqs. (11) and (13), together with intermediate states $|jm\rangle$ and $|j'm'\rangle$.

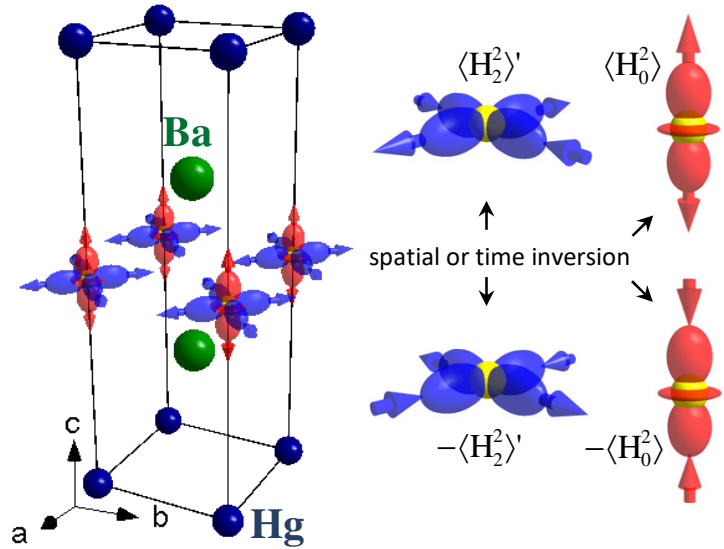

**Fig. 2**. Non-magnetic Hg1201 is tetragonal and Cu ions occupy sites that are centres of spatial inversion symmetry. However, emergence of time-reversal violation in the pseudo-gap phase drives a reduction in Cu site symmetry that includes the loss of inversion symmetry. The magnetic state is epitomized by the condensation of Dirac quadrupoles $\langle H^2_0 \rangle$ and $\langle H^2_{+2} \rangle'$ defined in Eq. (24) [38].

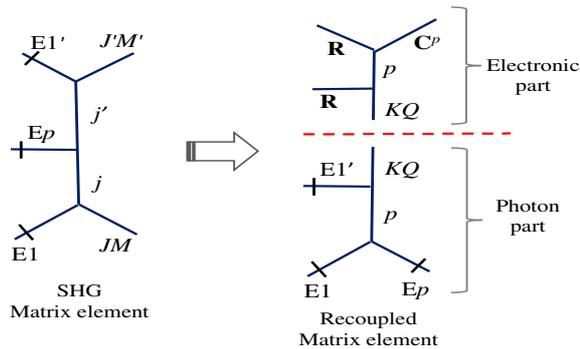

**Fig. 3**. Integrating out projections of intermediate states in the matrix element of the E1′-E$p$-E1 event creates an electronic spherical tensor operator with rank $K$, projection $Q$ and an RME given by Eq. (13). Photon and electronic tensors are linked by a mode label $p$, which has no analogy in two-photon events described by the Kramers-Heisenberg dispersion formula.

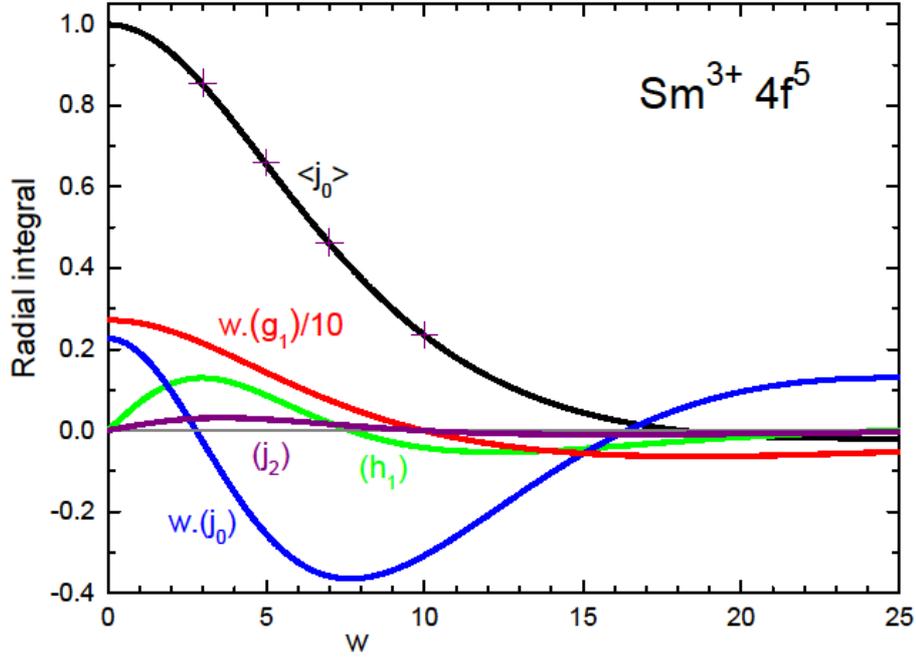

**Fig. 4**. Radial integrals for Dirac multipoles. Dimensionless variable $w = 12\pi a_o s$, where $a_o$ is the Bohr radius, while the standard variable for radial integrals $s$ is derived from the Bragg angle and neutron wavelength $s = \sin(\theta)/\lambda$. Legend: (——) $[w \times (g_1)]/10$, (——) $(h_1)$, (——) $[w \times (j_0)]$ and (——) $(j_2)$. Note that $(g_1)$ and $(j_0)$ from Eq. (A3) arise from the component of **Q** in Eq. (20) that contains the linear momentum operator and they are proportional to $1/w$ as the wavevector approaches zero. Atomic wavefunctions are $4f^5 - 5d^1$. Also included in the figure is the standard radial integral $\langle j_0(k) \rangle$. Results obtained with our $Sm^{3+}$ ($4f^5$) wavefunction are denoted by the continuous black curve, to which we added for comparison four values (**+**) derived from the standard interpolation formula [53].

---